\documentclass[doublecol]{epl2}

\usepackage{amsmath,amssymb,bm}
\usepackage{graphicx}
\usepackage{color}
\usepackage{url} 

\graphicspath{{figures/}}

\newcommand{\kb}{k_{\mathrm B}}
\newcommand{\dd}{\mathrm d}
\newcommand{\vv}{\bm v}
\newcommand{\rr}{\bm r}
\newcommand{\ff}{\bm F}
\newcommand{\rrand}{\bm R}

\newcommand{\GammaK}{\Gamma}

\title{DiffGLE: Differentiable Coarse-Grained Dynamics using Generalized Langevin Equation}
\shorttitle{Differentiable learning of memory kernels}

\author{Jinu Jeong\inst{1,*} \and Ishan Nadkarni\inst{2,*}}
\shortauthor{J. Jeong and I. Nadkarni}

\institute{
\inst{1} Department of Mechanical Science and Engineering, The University of Illinois at Urbana-Champaign, Urbana, Illinois, United States\\
\inst{2} Walker Department of Mechanical Engineering and Oden Institute for Computational Engineering and Sciences, The University of Texas at Austin, Austin, Texas, United States\\
\inst{*} These authors contributed equally to this work.
}


\abstract{
Capturing dynamical fidelity at the coarse-grained scale remains a central challenge in systematic molecular modelling. The generalized Langevin equation, rooted in the Mori--Zwanzig formalism, provides a principled framework for representing the memory friction and stochastic forces induced by eliminated microscopic degrees of freedom. In practice, however, parameterising its memory kernel is difficult because the exact kernel is a history-dependent projected-dynamics object coupled by the fluctuation--dissipation theorem to coloured random forces that are not directly observable from ordinary trajectories. Here we combine differentiable simulation with a coloured-noise ansatz to learn non-Markovian memory kernels in a top-down manner. The random force is represented by a trainable filter whose autocorrelation defines the friction memory, enforcing fluctuation--dissipation consistency by construction. The filter is optimised by backpropagating through coarse-grained generalized Langevin trajectories to match reference velocity autocorrelation functions, avoiding explicit projected-force reconstruction. We demonstrate the approach on bulk water, bulk carbon dioxide, and a single particle star-polymer memory benchmark. Across all these systems, the proposed framework accurately reproduces the target dynamical correlations, demonstrating that differentiable simulation enables direct optimisation of non-Markovian memory kernels from time-correlation observables.
}

\begin{document}

\maketitle

\section{Introduction}

Coarse-grained (CG) molecular models replace many atomistic degrees of freedom by a smaller set of effective variables. This reduction makes longer simulations possible, but it also changes the equations of motion. A conservative CG potential can often be fitted to reproduce equilibrium structural or thermodynamic observables through force matching, relative entropy, or iterative Boltzmann inversion \cite{noid_topdownbottom_2013,izvekov_multiscale_2005,chaimovich_coarse-graining_2011}. Dynamics are harder. Removing atoms, solvent, or internal modes removes both inertia and dissipation pathways, and CG simulations often move too quickly even when their structure appears satisfactory \cite{depa_why_2011,meinel_loss_2020,rudzinski_recent_2019}.

The Mori-Zwanzig projection formalism explains this discrepancy: exact reduced dynamics contain a conservative mean force, a history-dependent friction, and a fluctuating force \cite{mori_transport_1965,zwanzig_memory_1961,zwanzig_nonequilibrium_2001,hijon_morizwanzig_2009}. The resulting generalized Langevin equation (GLE) is formally appealing because the memory and random force are related by the fluctuation-dissipation theorem (FDT) \cite{kubo_fluctuation-dissipation_1966}. In practice, however, extracting or parameterising the memory kernel is difficult. Existing approaches include projected-force correlations, iterative reconstruction, auxiliary-variable embeddings, and direct memory-kernel optimisation \cite{izvekov_modeling_2006,li_incorporation_2015,lei_data-driven_2016,jung_iterative_2017,jung_generalized_2018,klippenstein_introducing_2021,klippenstein_gaussnewton_2024,del_razo_data-driven_2024}. Although these methods have substantially advanced non-Markovian coarse-grained modelling, they often require a dedicated reconstruction, embedding, or iterative parameterisation step before the resulting memory kernel can be used in simulation. We take a deliberately top-down alternative and formulate memory-kernel learning as a single differentiable inverse problem. The conservative CG force is kept fixed, while the unresolved dynamical contribution is represented by a trainable causal filter that generates the coloured random force. The filter autocorrelation defines the dissipative memory through the FDT, so stochastic forcing and friction remain consistent throughout optimisation. The resulting GLE simulator is then differentiated end-to-end, allowing the filter parameters to be trained directly from time-correlation observables such as the VACF. The key distinction is that the optimised object is not an independently reconstructed projected-force memory kernel, but an FDT-consistent coloured-noise generator whose induced memory is calibrated by the observable produced by the simulator itself. This work builds on differentiable simulation and neural differential equation ideas~\cite{chen_neural_2018,baydin_automatic_2018,wang_differentiable_2020,wang_learning_2023}, with the specific aim of learning non-Markovian friction kernels.

\begin{figure*}[t!]
\centering
\includegraphics[width=1.0\textwidth]{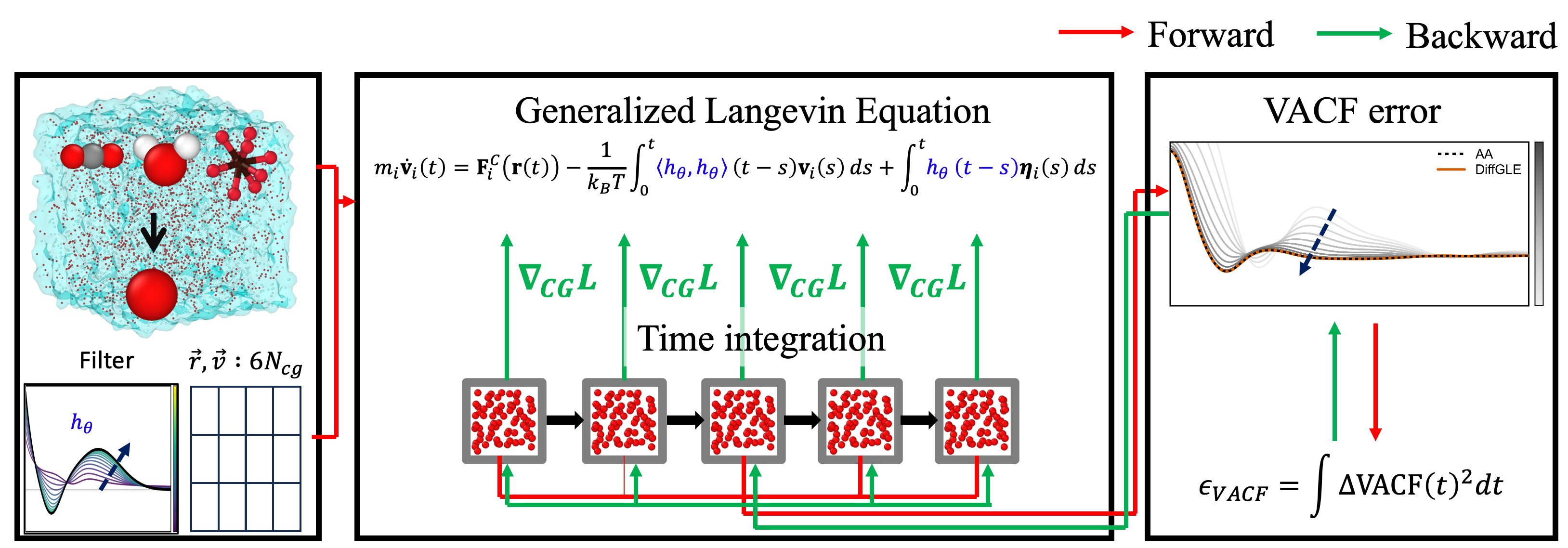}
\caption{
Schematic of the DiffGLE optimisation workflow. 
The CG state is propagated with a fixed conservative force and a trainable non-Markovian GLE memory kernel. 
The coloured-noise filter generates the random force, and its autocorrelation defines the memory kernel through the fluctuation--dissipation relation. 
Forward simulation produces a velocity autocorrelation function, while gradients of the VACF error are propagated backward through the differentiable trajectory to update the filter parameters.
}
\label{fig:methodology}
\end{figure*}

Figure~\ref{fig:methodology} summarises the resulting differentiable memory kernel optimization workflow.

\section{Differentiable GLE}

For CG coordinates $\rr_i$ and velocities $\vv_i$, we write the non-Markovian dynamics as
\begin{equation}
m_i \dot{\vv}_i(t)
=
\ff_i^{\mathrm C}(\rr(t))
-
\int_0^t \GammaK_\theta(t-s)\,\vv_i(s)\,\dd s
+
\rrand_{i,\theta}(t),
\label{eq:gle}
\end{equation}
where $\ff_i^{\mathrm C}$ is the fixed conservative CG force, $\GammaK_\theta$ is a trainable memory kernel, and $\rrand_{i,\theta}$ is a zero-mean fluctuating force. The second fluctuation-dissipation relation requires
\begin{equation}
\left\langle
R_{i\alpha,\theta}(t)R_{j\beta,\theta}(s)
\right\rangle
=
\kb T\,\GammaK_\theta(|t-s|)
\delta_{ij}\delta_{\alpha\beta}.
\label{eq:fdt}
\end{equation}
Rather than learning the fluctuating force and friction kernel independently,
we parameterise the random force through a causal coloured-noise filter,
\begin{equation}
R_{i,\theta}(t)
=
\int_0^\infty h_\theta(u)\eta_i(t-u)\,\mathrm du .
\label{eq:colored_noise_filter}
\end{equation}
Here $\eta_i$ is unit white noise with covariance
\begin{equation}
\left\langle
\eta_{i\alpha}(t)\eta_{j\beta}(s)
\right\rangle
=
\delta(t-s)\delta_{ij}\delta_{\alpha\beta}.
\label{eq:white_noise_covariance}
\end{equation}
The fluctuation--dissipation relation then fixes the memory kernel as the autocorrelation of this filter,
\begin{equation}
\Gamma_\theta(\tau)
=
\frac{1}{k_B T}
\int_0^\infty
h_\theta(u+\tau)h_\theta(u)\,\mathrm du .
\label{eq:filter_to_memory}
\end{equation}
In the discrete implementation, eq.~\eqref{eq:filter_to_memory} is evaluated as a finite autocorrelation of the learned filter.

\section{Gradient pathways for memory learning}

The optimisation problem above is an inverse problem through a stochastic, history-dependent simulator. It is therefore useful to make explicit what is being differentiated. After time discretisation, the GLE update can be written abstractly as
\begin{equation}
\bm x_{n+1}
=
\Phi_\theta(\bm x_n,\bm h_n,\bm \xi_n),
\label{eq:discrete_update}
\end{equation}
where $\bm x_n=(\bm r_n,\bm v_n)$ denotes the CG phase-space variables, $\bm h_n$ denotes the finite history required by the memory convolution, $\bm \xi_n$ is a fixed draw of the underlying white noise, and $\theta$ parameterises the coloured-noise filter. The history state contains the previous velocities and noise values needed to evaluate the dissipative and stochastic convolution terms. For a fixed random seed, eq. \eqref{eq:discrete_update} is a deterministic differentiable map from $\theta$ to the simulated trajectory and therefore to the VACF loss,
\begin{equation}
\mathcal L(\theta)
=
\ell\left(\bm x_0,\bm x_1,\ldots,\bm x_N;\theta\right).
\label{eq:trajectory_loss}
\end{equation}
This fixed-noise, or reparameterised, viewpoint is the basis
for all gradients used here: the optimiser differentiates the
response of the trajectory to the memory parameters, not
the probability of drawing a particular noise realisation
\cite{kingma_autoencoding_2014, rezende_stochastic_2014}.

The most direct gradient is obtained by differentiating the discrete simulator itself. Defining the trajectory sensitivity
\begin{equation}
\bm J_n = \frac{\partial \bm x_n}{\partial \theta},
\end{equation}
one obtains the forward sensitivity recursion
\begin{equation}
\bm J_{n+1}
=
\frac{\partial \Phi_\theta}{\partial \bm x_n}\bm J_n
+
\frac{\partial \Phi_\theta}{\partial \theta}.
\label{eq:forward_sensitivity}
\end{equation}
Equivalently, reverse-mode automatic differentiation through the same update gives
\begin{equation}
\bm a_n
=
\frac{\partial \ell}{\partial \bm x_n}
+
\left(
\frac{\partial \Phi_\theta}{\partial \bm x_n}
\right)^{\!T}
\bm a_{n+1},
\label{eq:discrete_adjoint}
\end{equation}
with parameter gradient
\begin{equation}
\nabla_\theta \mathcal L
=
\frac{\partial \ell}{\partial \theta}
+
\sum_{n=0}^{N-1}
\left(
\frac{\partial \Phi_\theta}{\partial \theta}
\right)^{\!T}
\bm a_{n+1}.
\label{eq:discrete_gradient}
\end{equation}
This discrete backpropagation route gives gradients of the actual numerical simulator. Its cost is that the reverse pass needs access to the forward trajectory and to the memory/noise history, so the memory footprint grows with trajectory length and kernel support.

For compact trajectories one can instead use an adjoint formulation. In continuous notation the augmented dynamics may be written as
\begin{equation}
\dot{\bm z}(t)
=
\bm f_\theta(\bm z(t),t;\bm \xi),
\label{eq:augmented_dynamics}
\end{equation}
where $\bm z$ contains positions, velocities, and any auxiliary variables needed to make the memory representation Markovian over the chosen state. For a loss
\begin{equation}
\mathcal L
=
\phi(\bm z(T))
+
\int_0^T q(\bm z(t),t;\theta)\,\mathrm dt,
\end{equation}
the adjoint variable
\begin{equation}
\bm a(t)=\frac{\partial \mathcal L}{\partial \bm z(t)}
\end{equation}
satisfies
\begin{equation}
\frac{\mathrm d\bm a}{\mathrm dt}
=
-
\left(
\frac{\partial \bm f_\theta}{\partial \bm z}
\right)^{\!T}
\bm a
-
\frac{\partial q}{\partial \bm z}.
\label{eq:continuous_adjoint}
\end{equation}
The parameter gradient is then
\begin{equation}
\nabla_\theta \mathcal L
=
\frac{\partial \phi}{\partial \theta}
+
\int_0^T
\left[
\frac{\partial q}{\partial \theta}
+
\left(
\frac{\partial \bm f_\theta}{\partial \theta}
\right)^{\!T}
\bm a(t)
\right]
\mathrm dt .
\label{eq:continuous_gradient}
\end{equation}
The advantage of this route is that the backward solve can avoid storing the entire forward computational graph. The important requirement is that the adjoint state must include all variables that influence future dynamics. For a Markovian ODE this is just the instantaneous state. For a non-Markovian GLE it also includes the representation of the memory and the fixed noise realisation, either explicitly through stored history or implicitly through auxiliary variables.

Both routes therefore compute gradients through the same physical pathway:
\begin{equation}
\theta
\longrightarrow
h_\theta
\longrightarrow
\Gamma_\theta
\longrightarrow
\{\bm r_n,\bm v_n\}_{n=0}^N
\longrightarrow
C_{vv}^{\mathrm{GLE}}(t)
\longrightarrow
\mathcal L .
\label{eq:gradient_pathway}
\end{equation}

In the differentiable GLE formulation, the trainable filter controls the dynamics through two coupled pathways. It generates the coloured random force and, through its autocorrelation, defines the FDT-consistent memory kernel. Changes in the filter therefore modify the stochastic forcing, the dissipative history term, the resulting velocity trajectory, and finally the VACF loss. The optimisation is thus performed on the observable produced by the GLE rollout itself, rather than on an intermediate projected-force correlation or separately reconstructed memory kernel.

\begin{figure*}[t!]
\centering
\includegraphics[width=0.82\textwidth]{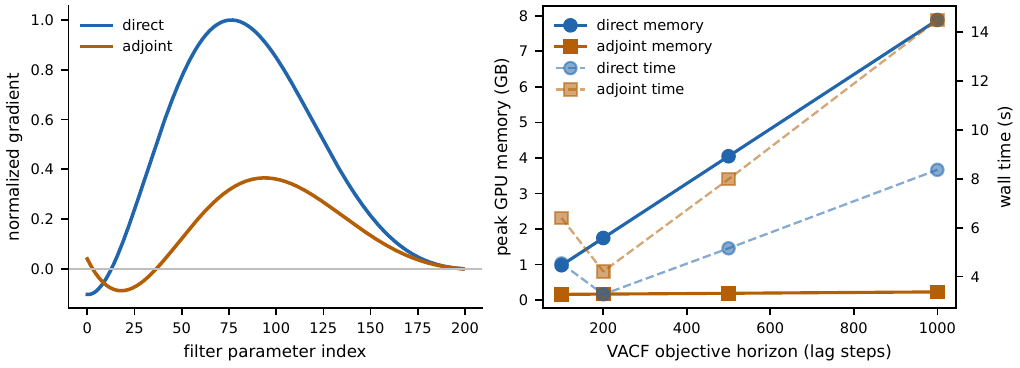}
\caption{
Gradient pathways and computational scaling for CO$_2$ memory learning.
Left: normalised gradient of the VACF loss with respect to the coloured-noise filter parameters, computed by direct backpropagation and by the adjoint route for the same fixed-noise trajectory.
Right: peak GPU memory and wall time as the VACF objective horizon is increased. Direct backpropagation stores the trajectory history and shows approximately linear memory growth, while the adjoint calculation keeps memory nearly constant by recomputing the trajectory during the backward pass.
}
\label{fig:gradient_mechanics}
\end{figure*}

The rollouts were implemented in a PyTorch-based molecular dynamics workflow. TorchMD's \texttt{System} object was used for molecular state, mass, and periodic-box bookkeeping, while the conservative forces, GLE thermostat, coloured-noise filter, VACF objective, and direct/adjoint differentiation were implemented as custom PyTorch/\texttt{torchdiffeq} modules~\cite{doerr_torchmd_2021,chen_neural_2018,baydin_automatic_2018,wang_differentiable_2020}. To assess the numerical consequences of the two differentiation routes, we performed a controlled gradient experiment on the CO$_2$ system using the same conservative force, filter parameters, random seed, forward trajectory, and VACF objective. Direct and adjoint differentiation therefore agree on the forward observable by construction; they differ only in how the derivative of that observable with respect to the filter is accumulated. Figure~\ref{fig:gradient_mechanics} summarises this comparison together with the corresponding computational-scaling measurement.

The filter-gradient profiles obtained by direct and adjoint differentiation have the same smooth, low-dimensional structure but differ in magnitude and detailed weighting across the filter support. This is expected for a history-dependent stochastic simulator unless the full memory state, noise realisation, and observable-history dependence are represented identically in the backward problem. The comparison therefore exposes where the learning signal enters the coloured-noise filter and how sensitive that signal is to the chosen differentiation pathway.

Second, the computational scaling displays the standard direct--adjoint tradeoff. Direct backpropagation stores the discrete trajectory and its memory history, so its peak GPU memory grows approximately linearly with the VACF objective horizon. The adjoint calculation recomputes the trajectory during the backward solve and therefore keeps the memory footprint nearly constant over the same range of horizons. This memory saving comes at the expected cost of additional wall time. Thus, direct differentiation is the most literal gradient of the implemented discrete simulator, whereas the adjoint route is the more scalable option when the augmented memory state is compact and accurately represented.

Both gradient strategies optimise the same FDT-consistent filter representation, but they are useful in different computational regimes. For the molecular-fluid systems, where the particle number is large and the learned memory is relatively compact, the adjoint-assisted formulation provides an efficient sensitivity calculation while retaining the coupling between coloured noise and friction through the filter autocorrelation. Direct fixed-noise differentiation, by contrast, keeps the full trajectory, noise realisation, and memory-convolution history inside the computational graph, so its cost grows rapidly with system size and memory support. This explicit pathwise-gradient formulation is nevertheless practical for the low-dimensional star-polymer benchmark considered here. In that setting it serves as a reference implementation in which the dependence of the VACF loss on the stochastic forcing history and the non-Markovian memory response is differentiated directly.
Therefore in practice, the choice between adjoint and direct differentiation is numerical rather than conceptual. The present results use both as differentiable estimators of the same objective: learning an FDT-consistent memory kernel that makes the CG VACF match the reference dynamics.

\section{Results}
\begin{figure*}[t!]
\centering
\includegraphics[width=\textwidth]{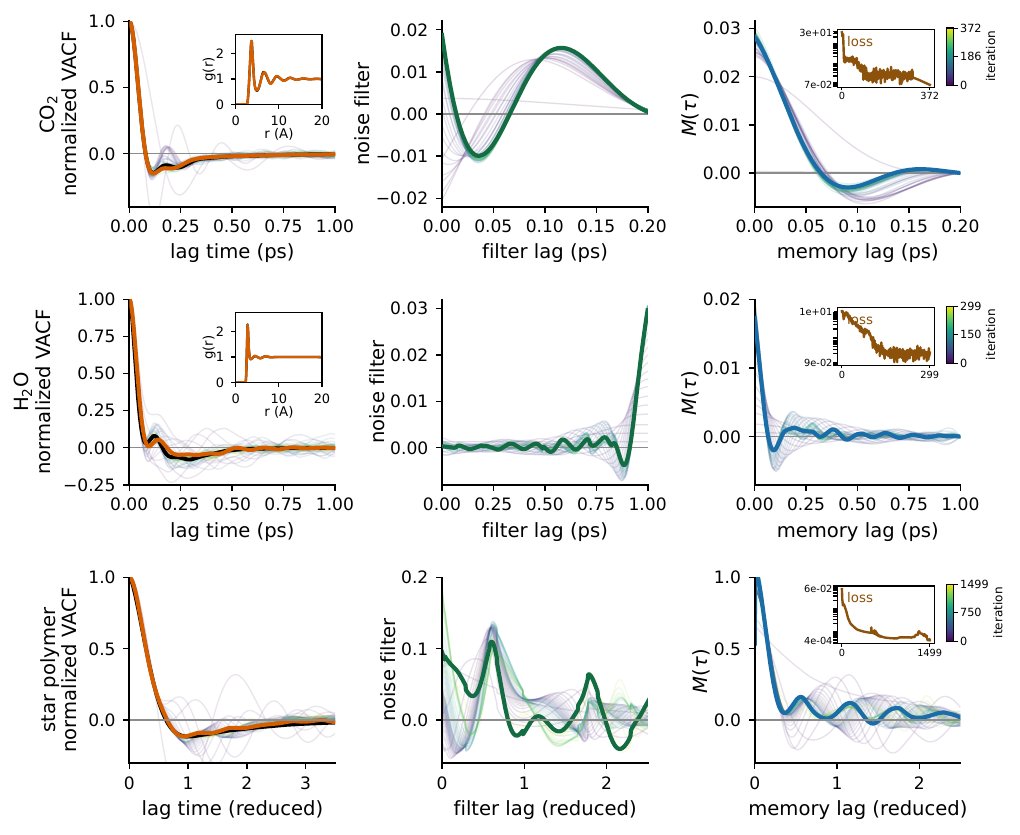}
\caption{
Differentiable memory learning across the three benchmark systems. Rows show CO$_2$, H$_2$O, and the force-free star-polymer memory benchmark. In each row, the left panel compares the reference VACF, intermediate differentiable-GLE trajectories, and the final DiffGLE result; black denotes the reference target and orange denotes the final DiffGLE trajectory. RDF insets are included for the molecular fluids. The middle and right panels show the learned coloured-noise filter and the FDT memory kernel computed from its autocorrelation; in these panels $M(\tau)\equiv\Gamma_\theta(\tau)$. Thick green and blue curves denote the final optimised filter and memory kernel, while faint curves show intermediate optimisation iterates, not statistical uncertainty bands. Compact colourbars indicate optimisation iteration, and small insets in the memory-kernel panels show the training loss. Time axes are in ps for the molecular fluids and in reduced benchmark units for the star-polymer row.
}
\label{fig:main_results}
\end{figure*}
\subsection{Carbon dioxide}

Carbon dioxide provides a compact demonstration of the full adjoint-assisted learning workflow. Starting from an initially unstructured filter, stochastic-gradient optimisation progressively modifies the coloured-noise response and the corresponding FDT-induced memory kernel until the simulated VACF approaches the all-atom reference. The optimised result shown in the top row of fig.~\ref{fig:main_results} is obtained after a short continuation stage used to stabilise the final filter. Production validation gives a VACF RMSE of $6.98\times 10^{-3}$ and an RDF RMSE of $1.02\times 10^{-2}$, indicating that the learned dynamical correction improves time correlations while retaining the equilibrium pair structure.

\subsection{Water}

Water presents a more stringent molecular-fluid test because its VACF contains a rapid inertial decay followed by a weak oscillatory tail. The middle row of fig.~\ref{fig:main_results} shows the optimisation trajectory from the adjoint-assisted training run, in which the learned kernel reaches a training RMSE of approximately $1.5\times 10^{-2}$. The optimisation first corrects the short-time decay and then adjusts the small negative and oscillatory portions of the VACF, indicating that the learned memory is not acting only as an effective scalar damping. Instead, the final kernel contains a dominant short-time friction component together with weaker sign-changing contributions that tune the phase and amplitude of the tail.

The combined molecular-fluid rows emphasise the intended separation of roles. The fixed conservative interaction keeps the RDF close to the reference, while the learned filter changes the stochastic and dissipative terms that control the VACF. The two fluids require different memory shapes, but the optimisation pathway is the same in both cases.

\subsection{Star-polymer memory benchmark}

The third test decouples memory learning from conservative-force errors by considering a force-free star-polymer benchmark motivated by non-Markovian coarse-grained descriptions of soft-matter dynamics~\cite{li_incorporation_2015}. The reference VACF is generated from our TorchMD implementation of the Li et al. star-polymer benchmark with the conservative force disabled, so the comparison isolates the memory contribution. The reference dynamics exhibit a slow VACF relaxation that cannot be represented accurately by a single optimised Markovian friction coefficient. We therefore train a long-support filter using fixed-noise direct backpropagation. The final DiffGLE model, shown in the bottom row of fig.~\ref{fig:main_results}, achieves a VACF RMSE of $1.98\times 10^{-2}$, compared with $7.60\times 10^{-2}$ for the optimised Markovian baseline. This result shows that the filter representation can learn a broad memory response when the target dynamics require frequency-dependent friction beyond a single timescale.

Together, the benchmarks probe short molecular memory in CO$_2$, sharper inertial decay with a weak tail in water, and the long-memory limit in the force-free star-polymer case. In all three, the conservative model is fixed; optimisation acts only through the coloured-noise filter and its FDT-consistent kernel, with structural quantities used as guardrails. The final production VACF RMSEs are $6.98\times10^{-3}$ for CO$_2$, $2.18\times10^{-2}$ for H$_2$O, and $1.98\times10^{-2}$ for the star polymer; the molecular-fluid RDF RMSEs are $1.02\times10^{-2}$ and $3.37\times10^{-3}$.


\section{Conclusion and outlook}

We have introduced a top-down differentiable framework for parameterising the non-Markovian memory kernel of coarse-grained generalized Langevin dynamics. The conservative coarse-grained interaction is kept fixed, whereas the missing dynamical contribution is learned through a colored-noise filter whose autocorrelation defines the friction memory. This construction enforces the equilibrium fluctuation--dissipation relation between stochastic forcing and dissipation by design, and converts memory-kernel parameterization into an optimization problem over filter coefficients. By differentiating through coarse-grained trajectories, the filter can be trained directly against reference time-correlation observables. The resulting models improve velocity-autocorrelation agreement for molecular fluids while preserving structural consistency, and capture slow memory relaxation in a star-polymer benchmark beyond what can be represented by an optimized Markovian friction.

The practical implication is that memory learning can be treated as a dynamical calibration layer on top of an existing coarse-grained model. This separation is useful when a conservative potential has already been validated structurally, or when retraining it would obscure whether dynamical errors arise from equilibrium thermodynamics or from missing frictional memory. In that setting, the VACF or another time-correlation observable provides a direct target for the non-Markovian part of the dynamics.

The present examples are intentionally restricted to fixed conservative forces and equilibrium time-correlation targets. This makes the role of the learned memory kernel transparent, but it also leaves several natural extensions open. In molecular applications with anisotropic or state-dependent friction, the scalar filter used here would need to be replaced by a matrix-valued or local environment-dependent noise generator. For observables with long relaxation times, variance reduction and checkpointed differentiation become as important as the kernel parameterisation itself. These limitations are numerical rather than conceptual: the FDT construction still supplies the link between coloured noise and memory, while the differentiable simulator determines which observable is optimized.

The same paradigm suggests a route toward driven and nonequilibrium coarse-grained systems, where the optimization target may be a steady-state response rather than an equilibrium VACF. Field-dependent mobility, confined-flow velocity profiles, shear or mass fluxes, ionic currents, and transport coefficients could all define losses for learning state- or geometry-dependent dynamical kernels. Such extensions require care because the equilibrium FDT used here need not hold unchanged far from equilibrium. Recent work on nonequilibrium fluctuation--dissipation relations, non-stationary GLEs, and data-driven GLE parameterisation provides a foundation for this direction~\cite{zhu_generalized_2021, cui_external_2018,netz_nonequilibrium_2024,meyer_dynamics_2019,meyer_nonmarkovian_2020,glatzel_validity_2021,xie_abinitio_2022,xie_multidimensional_2024}. The central idea remains the same: dynamical fidelity can be optimised directly from physically meaningful time-dependent observables.

\section {Data Availability Statement}
The code, preprocessed inputs, and analysis scripts required to reproduce the reported figures are available at \url{https://github.com/jinu-jeong/Differentiable_GLE}. Additional simulation parameters, optimisation details, and validation plots are included in the Supplementary Material.

\acknowledgments

The work on deep learning was supported by the Center for Enhanced Nanofluidic Transport (CENT), an Energy Frontier Research Center funded by the U.S. Department of Energy, Office of Science, Basic Energy Sciences (Award No. DE-SC0019112). All other aspects of this work were supported by the National Science Foundation under Grant No 2137157. The authors acknowledge the Texas Advanced Computing Center (TACC) at The University of Texas at Austin for providing access to the Lonestar6 resource that has contributed to the research results reported within this paper. We also acknowledge the use of the Extreme Science and Engineering Discovery Environment (XSEDE) Stampede2 at the Texas Advanced Computing Centre through Allocation No. TG-CDA100010.

\bibliographystyle{unsrt}
\bibliography{references}

@article{mori_transport_1965,
	title = {Transport, {Collective} {Motion}, and {Brownian} {Motion}*)},
	volume = {33},
	issn = {0033-068X},
	url = {https://doi.org/10.1143/PTP.33.423},
	doi = {10.1143/PTP.33.423},
	number = {3},
	urldate = {2023-04-24},
	journal = {Progress of Theoretical Physics},
	author = {Mori, Hazime},
	month = mar,
	year = {1965},
	pages = {423--455},
}

@article{zwanzig_memory_1961,
	title = {Memory {Effects} in {Irreversible} {Thermodynamics}},
	volume = {124},
	issn = {0031-899X},
	url = {https://link.aps.org/doi/10.1103/PhysRev.124.983},
	doi = {10.1103/PhysRev.124.983},
	language = {en},
	number = {4},
	urldate = {2023-11-20},
	journal = {Physical Review},
	author = {Zwanzig, Robert},
	month = nov,
	year = {1961},
	pages = {983--992},
}

@article{kubo_fluctuation-dissipation_1966,
	title = {The fluctuation-dissipation theorem},
	volume = {29},
	issn = {0034-4885},
	url = {https://dx.doi.org/10.1088/0034-4885/29/1/306},
	doi = {10.1088/0034-4885/29/1/306},
	language = {en},
	number = {1},
	urldate = {2023-11-27},
	journal = {Reports on Progress in Physics},
	author = {Kubo, R},
	month = jan,
	year = {1966},
	pages = {255},
}

@article{hijon_morizwanzig_2009,
	title = {Mori–{Zwanzig} formalism as a practical computational tool},
	volume = {144},
	issn = {1364-5498},
	url = {https://pubs.rsc.org/en/content/articlelanding/2010/fd/b902479b},
	doi = {10.1039/B902479B},
	language = {en},
	number = {0},
	urldate = {2023-04-12},
	journal = {Faraday Discussions},
	publisher = {The Royal Society of Chemistry},
	author = {Hijón, Carmen and Español, Pep and Vanden-Eijnden, Eric and Delgado-Buscalioni, Rafael},
	month = oct,
	year = {2009},
	pages = {301--322},
}

@article{noid_topdownbottom_2013,
	title = {{TopDown}/{Bottom} up: {Perspective}: {Coarse}-grained models for biomolecular systems},
	volume = {139},
	issn = {0021-9606},
	shorttitle = {Perspective},
	url = {https://aip.scitation.org/doi/10.1063/1.4818908},
	doi = {10.1063/1.4818908},
	number = {9},
	urldate = {2023-01-13},
	journal = {The Journal of Chemical Physics},
	publisher = {American Institute of Physics},
	author = {Noid, W. G.},
	month = sep,
	year = {2013},
	pages = {090901},
}

@article{izvekov_multiscale_2005,
	title = {A {Multiscale} {Coarse}-{Graining} {Method} for {Biomolecular} {Systems}},
	volume = {109},
	issn = {1520-6106},
	url = {https://doi.org/10.1021/jp044629q},
	doi = {10.1021/jp044629q},
	number = {7},
	urldate = {2023-04-12},
	journal = {The Journal of Physical Chemistry B},
	publisher = {American Chemical Society},
	author = {Izvekov, Sergei and Voth, Gregory A.},
	month = feb,
	year = {2005},
	pages = {2469--2473},
}

@article{chaimovich_coarse-graining_2011,
	title = {Coarse-graining errors and numerical optimization using a relative entropy framework},
	volume = {134},
	issn = {0021-9606, 1089-7690},
	url = {http://aip.scitation.org/doi/10.1063/1.3557038},
	doi = {10.1063/1.3557038},
	language = {en},
	number = {9},
	urldate = {2022-11-14},
	journal = {The Journal of Chemical Physics},
	author = {Chaimovich, Aviel and Shell, M. Scott},
	month = mar,
	year = {2011},
	pages = {094112},
}

@article{rudzinski_recent_2019,
	title = {Recent {Progress} towards {Chemically}-{Specific} {Coarse}-{Grained} {Simulation} {Models} with {Consistent} {Dynamical} {Properties}},
	volume = {7},
	copyright = {http://creativecommons.org/licenses/by/3.0/},
	issn = {2079-3197},
	url = {https://www.mdpi.com/2079-3197/7/3/42},
	doi = {10.3390/computation7030042},
	language = {en},
	number = {3},
	urldate = {2023-04-16},
	journal = {Computation},
	publisher = {Multidisciplinary Digital Publishing Institute},
	author = {Rudzinski, Joseph F.},
	month = sep,
	year = {2019},
	note = {Number: 3},
	pages = {42},
}

@article{izvekov_modeling_2006,
	title = {Modeling real dynamics in the coarse-grained representation of condensed phase systems},
	volume = {125},
	issn = {0021-9606, 1089-7690},
	url = {https://pubs.aip.org/jcp/article/125/15/151101/937909/Modeling-real-dynamics-in-the-coarse-grained},
	doi = {10.1063/1.2360580},
	language = {en},
	number = {15},
	urldate = {2023-11-20},
	journal = {The Journal of Chemical Physics},
	author = {Izvekov, Sergei and Voth, Gregory A.},
	month = oct,
	year = {2006},
	pages = {151101},
}

@article{depa_why_2011,
	title = {Why are coarse-grained force fields too fast? {A} look at dynamics of four coarse-grained polymers},
	volume = {134},
	issn = {0021-9606, 1089-7690},
	shorttitle = {Why are coarse-grained force fields too fast?},
	url = {https://pubs.aip.org/jcp/article/134/1/014903/564687/Why-are-coarse-grained-force-fields-too-fast-A},
	doi = {10.1063/1.3513365},
	language = {en},
	number = {1},
	urldate = {2023-11-27},
	journal = {The Journal of Chemical Physics},
	author = {Depa, Praveen and Chen, Chunxia and Maranas, Janna K.},
	month = jan,
	year = {2011},
	pages = {014903},
}

@article{meinel_loss_2020,
	title = {Loss of {Molecular} {Roughness} upon {Coarse}-{Graining} {Predicts} the {Artificially} {Accelerated} {Mobility} of {Coarse}-{Grained} {Molecular} {Simulation} {Models}},
	volume = {16},
	issn = {1549-9618, 1549-9626},
	url = {https://pubs.acs.org/doi/10.1021/acs.jctc.9b00943},
	doi = {10.1021/acs.jctc.9b00943},
	language = {en},
	number = {3},
	urldate = {2023-11-22},
	journal = {Journal of Chemical Theory and Computation},
	author = {Meinel, Melissa K. and Müller-Plathe, Florian},
	month = mar,
	year = {2020},
	pages = {1411--1419},
}

@article{li_incorporation_2015,
	title = {Incorporation of memory effects in coarse-grained modeling via the {Mori}-{Zwanzig} formalism},
	volume = {143},
	issn = {0021-9606, 1089-7690},
	url = {https://pubs.aip.org/jcp/article/143/24/243128/963573/Incorporation-of-memory-effects-in-coarse-grained},
	doi = {10.1063/1.4935490},
	language = {en},
	number = {24},
	urldate = {2023-11-21},
	journal = {The Journal of Chemical Physics},
	author = {Li, Zhen and Bian, Xin and Li, Xiantao and Karniadakis, George Em},
	month = dec,
	year = {2015},
	pages = {243128},
}

@article{jung_iterative_2017,
	title = {Iterative {Reconstruction} of {Memory} {Kernels}},
	volume = {13},
	issn = {1549-9618},
	url = {https://doi.org/10.1021/acs.jctc.7b00274},
	doi = {10.1021/acs.jctc.7b00274},
	number = {6},
	urldate = {2023-11-21},
	journal = {Journal of Chemical Theory and Computation},
	publisher = {American Chemical Society},
	author = {Jung, Gerhard and Hanke, Martin and Schmid, Friederike},
	month = jun,
	year = {2017},
	pages = {2481--2488},
}

@article{jung_generalized_2018,
	title = {Generalized {Langevin} dynamics: construction and numerical integration of non-{Markovian} particle-based models},
	volume = {14},
	issn = {1744-6848},
	shorttitle = {Generalized {Langevin} dynamics},
	url = {https://pubs.rsc.org/en/content/articlelanding/2018/sm/c8sm01817k},
	doi = {10.1039/C8SM01817K},
	language = {en},
	number = {46},
	urldate = {2023-11-21},
	journal = {Soft Matter},
	publisher = {The Royal Society of Chemistry},
	author = {Jung, Gerhard and Hanke, Martin and Schmid, Friederike},
	month = nov,
	year = {2018},
	pages = {9368--9382},
}

@article{klippenstein_introducing_2021,
	title = {Introducing {Memory} in {Coarse}-{Grained} {Molecular} {Simulations}},
	volume = {125},
	issn = {1520-6106, 1520-5207},
	url = {https://pubs.acs.org/doi/10.1021/acs.jpcb.1c01120},
	doi = {10.1021/acs.jpcb.1c01120},
	language = {en},
	number = {19},
	urldate = {2023-09-03},
	journal = {The Journal of Physical Chemistry B},
	author = {Klippenstein, Viktor and Tripathy, Madhusmita and Jung, Gerhard and Schmid, Friederike and Van Der Vegt, Nico F. A.},
	month = may,
	year = {2021},
	pages = {4931--4954},
}

@article{klippenstein_gaussnewton_2024,
	title = {A {Gauss}–{Newton} method for iterative optimization of memory kernels for generalized {Langevin} thermostats in coarse-grained molecular dynamics simulations},
	volume = {160},
	issn = {0021-9606},
	url = {https://doi.org/10.1063/5.0203832},
	doi = {10.1063/5.0203832},
	number = {20},
	urldate = {2024-09-18},
	journal = {The Journal of Chemical Physics},
	author = {Klippenstein, Viktor and Wolf, Niklas and van der Vegt, Nico F. A.},
	month = may,
	year = {2024},
	pages = {204115},
}

@article{del_razo_data-driven_2024,
	title = {Data-driven dynamical coarse-graining for condensed matter systems},
	volume = {160},
	issn = {0021-9606},
	url = {https://doi.org/10.1063/5.0177553},
	doi = {10.1063/5.0177553},
	number = {2},
	urldate = {2024-07-09},
	journal = {The Journal of Chemical Physics},
	author = {del Razo, Mauricio J. and Crommelin, Daan and Bolhuis, Peter G.},
	month = jan,
	year = {2024},
	pages = {024108},
}

@inproceedings{chen_neural_2018,
  author    = {Chen, Ricky T. Q. and Rubanova, Yulia and Bettencourt, Jesse and Duvenaud, David K.},
  title     = {Neural Ordinary Differential Equations},
  booktitle = {Advances in Neural Information Processing Systems 31},
  editor    = {Bengio, S. and Wallach, H. and Larochelle, H. and Grauman, K. and Cesa-Bianchi, N. and Garnett, R.},
  pages     = {6571--6583},
  publisher = {Curran Associates, Inc.},
  year      = {2018},
  eprint    = {1806.07366},
  archivePrefix = {arXiv},
  primaryClass  = {cs.LG}
}

@article{baydin_automatic_2018,
  author  = {Baydin, Atilim Gunes and Pearlmutter, Barak A. and Radul, Alexey Andreyevich and Siskind, Jeffrey Mark},
  title   = {Automatic Differentiation in Machine Learning: a Survey},
  journal = {Journal of Machine Learning Research},
  volume  = {18},
  number  = {153},
  pages   = {1--43},
  year    = {2018},
  url     = {http://jmlr.org/papers/v18/17-468.html}
}

@misc{wang_differentiable_2020,
	title = {Differentiable {Molecular} {Simulations} for {Control} and {Learning}},
	url = {http://arxiv.org/abs/2003.00868},
	doi = {10.48550/arXiv.2003.00868},
	urldate = {2023-12-24},
	publisher = {arXiv},
	author = {Wang, Wujie and Axelrod, Simon and Gómez-Bombarelli, Rafael},
	month = dec,
	year = {2020},
	note = {arXiv:2003.00868 [physics, stat]},
}

@article{wang_learning_2023,
	title = {Learning pair potentials using differentiable simulations},
	volume = {158},
	issn = {0021-9606},
	url = {https://doi.org/10.1063/5.0126475},
	doi = {10.1063/5.0126475},
	number = {4},
	urldate = {2023-12-24},
	journal = {The Journal of Chemical Physics},
	author = {Wang, Wujie and Wu, Zhenghao and Dietschreit, Johannes C. B. and Gómez-Bombarelli, Rafael},
	month = jan,
	year = {2023},
	pages = {044113},
}

@book{zwanzig_nonequilibrium_2001,
  title = {Nonequilibrium Statistical Mechanics},
  author = {Zwanzig, Robert},
  year = {2001},
  publisher = {Oxford University Press},
  address = {Oxford}
}

@article{lei_data-driven_2016,
  title = {Data-driven parameterization of the generalized Langevin equation},
  author = {Lei, Huan and Baker, Nathan A. and Li, Xiantao},
  journal = {Proceedings of the National Academy of Sciences of the United States of America},
  volume = {113},
  number = {50},
  pages = {14183--14188},
  year = {2016},
  doi = {10.1073/pnas.1609587113}
}

@article{doerr_torchmd_2021,
  title = {TorchMD: A Deep Learning Framework for Molecular Simulations},
  author = {Doerr, Stefan and Majewski, Maciej and P{\'e}rez, Adri{\`a} and Kr{\"a}mer, Andreas and Clementi, Cecilia and No{\'e}, Frank and Giorgino, Toni and De Fabritiis, Gianni},
  journal = {Journal of Chemical Theory and Computation},
  volume = {17},
  number = {4},
  pages = {2355--2363},
  year = {2021},
  doi = {10.1021/acs.jctc.0c01343}
}

@article{zhu_generalized_2021,
  author  = {Zhu, Yuanran and Lei, Huan and Kim, Changho},
  title   = {Generalized second fluctuation-dissipation theorem in nonequilibrium steady states},
  journal = {Phys. Scr.},
  volume  = {98},
  pages   = {115402},
  year    = {2023},
  doi     = {10.1088/1402-4896/acfce5}
}

@article{cui_external_2018,
  author  = {Cui, Bingyu and Zaccone, Alessio},
  title   = {Generalized Langevin equation and fluctuation-dissipation theorem for particle-bath systems in external oscillating fields},
  journal = {Physical Review E},
  volume  = {97},
  pages   = {060102},
  year    = {2018},
  doi     = {10.1103/PhysRevE.97.060102}
}

@article{netz_nonequilibrium_2024,
  author  = {Netz, Roland R.},
  title   = {Non-equilibrium generalized Langevin equation from a time-dependent Hamiltonian},
  journal = {Phys. Rev. E},
  volume  = {110},
  pages   = {014123},
  year    = {2024},
  doi     = {10.1103/PhysRevE.110.014123}
}

@article{meyer_dynamics_2019,
  author  = {Meyer, Hugues and Voigtmann, Thomas and Schilling, Tanja},
  title   = {Dynamics of reaction coordinates in time-dependent many-body processes},
  journal = {J. Chem. Phys.},
  volume  = {150},
  pages   = {174118},
  year    = {2019},
  doi     = {10.1063/1.5090450}
}

@article{meyer_nonmarkovian_2020,
  author  = {Meyer, Hugues and Pelagejcev, Philipp and Schilling, Tanja},
  title   = {Non-Markovian out-of-equilibrium dynamics and time-dependent memory kernels},
  journal = {EPL},
  volume  = {128},
  pages   = {40001},
  year    = {2020},
  doi     = {10.1209/0295-5075/128/40001}
}

@article{glatzel_validity_2021,
  author  = {Glatzel, Fabian and Schilling, Tanja},
  title   = {Validity of the non-stationary generalized Langevin equation for coarse-grained stochastic dynamics},
  journal = {J. Chem. Phys.},
  volume  = {154},
  pages   = {174107},
  year    = {2021},
  doi     = {10.1063/5.0049693}
}

@article{xie_multidimensional_2024,
  author  = {Xie, Pinchen and Qiu, Yunrui and E, Weinan},
  title   = {Coarse-graining conformational dynamics with multidimensional generalized Langevin equations},
  journal = {J. Chem. Theory Comput.},
  year    = {2024},
  doi     = {10.1021/acs.jctc.4c00729}
}

@article{xie_abinitio_2022,
  author        = {Xie, Pinchen and Car, Roberto and E, Weinan},
  title         = {Ab initio generalized Langevin equation},
  journal       = {Proceedings of the National Academy of Sciences},
  volume        = {121},
  pages         = {e2308668121},
  year          = {2024},
  doi           = {10.1073/pnas.2308668121},
  eprint        = {2211.06558},
  archivePrefix = {arXiv},
  primaryClass  = {physics.comp-ph}
}

@inproceedings{kingma_autoencoding_2014,
  author    = {Kingma, Diederik P. and Welling, Max},
  title     = {Auto-Encoding Variational Bayes},
  booktitle = {International Conference on Learning Representations},
  year      = {2014},
  eprint    = {1312.6114},
  archivePrefix = {arXiv},
  primaryClass  = {stat.ML}
}

@inproceedings{rezende_stochastic_2014,
  author    = {Rezende, Danilo Jimenez and Mohamed, Shakir and Wierstra, Daan},
  title     = {Stochastic Backpropagation and Approximate Inference in Deep Generative Models},
  booktitle = {Proceedings of the 31st International Conference on Machine Learning},
  series    = {Proceedings of Machine Learning Research},
  volume    = {32},
  pages     = {1278--1286},
  year      = {2014},
  eprint    = {1401.4082},
  archivePrefix = {arXiv},
  primaryClass  = {stat.ML}
}

\end{document}